\documentclass[journal]{IEEEtran}

\usepackage{amsmath,amssymb,graphicx}
\usepackage{wrapfig}
\usepackage{tabularx}
\usepackage{booktabs}
\usepackage{microtype}
\usepackage{textcomp}
\usepackage{makecell}
\usepackage{framed}
\usepackage{dblfloatfix}
\usepackage[separate-uncertainty=true, multi-part-units=single]{siunitx}
\usepackage{bm}
\usepackage[hidelinks]{hyperref}
\usepackage{enumitem}
\usepackage{setspace}
\usepackage{color}
\usepackage{wrapfig}
\usepackage{float}
\usepackage{soulutf8}
\usepackage{filecontents}
\usepackage{etoolbox}
\newcolumntype{B}{<{\hspace{-1ex}}c}
\usepackage{multirow}
\usepackage{glossaries}
\makeglossaries
\DeclareSIUnit\px{pixels}
\DeclareSIUnit\db{dB}
\newcolumntype{Y}{>{\centering\arraybackslash}X}

\usepackage[style=ieee,backend=biber, url=false, doi=false, isbn=false, eprint=false,maxbibnames=6,minbibnames=1]{biblatex}
\AtEveryBibitem{%
	\clearfield{day}%
	\clearfield{month}%
	\clearfield{endday}%
	\clearfield{endmonth}%
	\clearlist{language}%
	\clearfield{country}
	\clearfield{city}
}

\newcommand{\irow}[1]{
	\begin{matrix}(#1)\end{matrix}%
}
\newcommand{\ignore}[1]{}
\expandafter\def\expandafter\normalsize\expandafter{%
	\normalsize
	\setlength\abovedisplayskip{6pt}
	\setlength\belowdisplayskip{6pt}
	\setlength\abovedisplayshortskip{6pt}
	\setlength\belowdisplayshortskip{6pt}
}

\newacronym{psf}{PSF}{Point Spread Function}
\newacronym{snr}{SNR}{Signal-to-Noise Ratio}
\newacronym{bd}{BD}{Blind Deconvolution}
\newacronym{cnn}{CNN}{Convolutional Neural Network}
\newacronym{map}{MAP}{Maximum a Posteriori}
\newacronym{tv}{TV}{Total Variation}
\newacronym{tvrl}{TV-RL}{Total Variation regularized Richardson-Lucy}
\newacronym{rl}{RL}{Richardson-Lucy}
\newacronym{ssim}{SSIM}{Structural Similarity}

\begin{document}

\title{{\vspace{-0cm}Spatially-Variant CNN-based Point Spread Function Estimation for 
Blind Deconvolution and Depth Estimation
in Optical Microscopy}}
\author{Adrian~Shajkofci,~\IEEEmembership{Graduate Student Member,~IEEE,}
	Michael~Liebling,~\IEEEmembership{Member,~IEEE}%
\thanks{Adrian Shajkofci is with the Idiap Research Institute, CH-1920 Martigny, Switzerland and \'Ecole Polytechnique F\'ed\'erale de Lausanne, CH-1015 Lausanne, Switzerland.}%
\thanks{Michael Liebling is with the Idiap Research Institute and the Electrical \& Computer Engineering Department, University of California, Santa Barbara, CA 93106, USA.}%
\thanks{Manuscript received December 3, 2019; revised March 3, 2020; accepted April 2, 2020. Date of publication April 15, 2020; date of current version April 27, 2020. This work was supported by the Swiss National Science Foundation under Grant 206021\_164022 and Grant 200020\_179217. The associate editor coordinating the review of this manuscript and approving itfor publication was Prof. Dong Xu.}
\thanks{Digital Object Identifier 10.1109/TIP.2020.2986880}
\thanks{Copyright 2020 IEEE. Personal use of this material is permitted. Permission from IEEE must be obtained for all other uses, including reprinting/republishing this material for advertising or promotional purposes, collecting new collected works for resale or redistribution to servers or lists, or reuse of any copyrighted component of this work in other works.}}
\markboth{IEEE TRANSACTIONS ON IMAGE PROCESSING, VOL. 29, 2020}
{Shajkofci \MakeLowercase{\textit{et al.}}: Spatially-Variant CNN-based Point Spread Function Estimation for 2.5D Optical Microscopy Reconstruction via Blind Deconvolution and Depth Estimation}

\maketitle

\begin{abstract}
Optical microscopy is an essential tool in biology and medicine. Imaging thin, yet non-flat objects in a single shot (without relying on more sophisticated sectioning setups) remains challenging as the shallow depth of field that comes with high-resolution microscopes leads to unsharp image regions and makes depth localization and quantitative image interpretation difficult. 

Here, we present a method that improves the resolution of light microscopy images of such objects by locally estimating image distortion while jointly estimating object distance to the focal plane. Specifically, we estimate the parameters of a spatially-variant \gls{psf} model using a \gls{cnn}, which does not require instrument- or object-specific calibration. Our method recovers \gls{psf} parameters from the image itself with up to a squared Pearson correlation coefficient of 0.99 in ideal conditions, while remaining robust to object rotation, illumination variations, or photon noise. When the recovered \glspl{psf} are used with a spatially-variant and regularized \gls{rl} deconvolution algorithm, we observed up to 2.1 dB better \gls{snr} compared to other \gls{bd} techniques. Following microscope-specific calibration, we further demonstrate that the recovered \gls{psf} model parameters permit  estimating surface depth with a precision of 2 micrometers and over an extended range when using engineered \glspl{psf}. Our method opens up multiple possibilities for enhancing images of non-flat objects with minimal need for a priori knowledge about the optical setup.
\end{abstract}


\section{Introduction}
\label{sec:intro}
\glsresetall
\IEEEPARstart{R}{esearchers} and physicians intensively use optical microscopes to observe and quantify cellular function, organ development, or disease mechanisms.
Despite the availability of many volumetric imaging methods (in particular, optical sectioning methods), single-shot wide-field microscopy remains an important tool to image small  and relatively shallow objects. However, non-flat areas, which are out of focus, lead to unsharp regions in the image, making localization and visual interpretation difficult. 
Image formation in a microscope can be modeled by light diffraction, which causes sharp point-like objects to appear blurry \cite{gibson_diffraction_1989}. Because the optical system only collects a fraction of the light emanating from a point on the object, it cannot focus the light into a perfect point and, instead, spreads the light into a three-dimensional diffraction pattern described by the \gls{psf}. As the image is formed by superposing the contribution of all points in the object, knowledge of the local diffraction pattern, which sums up the optical system and its aberrations, can be used to estimate a sharper image \cite{sage_deconvolutionlab2_2017}. 

For thin, yet not flat samples, image formation can be modeled as a superposition of 2D \glspl{psf}. These are shaped both by the optical system and the three-dimensional depth of the object. 
Knowledge of the local \gls{psf} could therefore both be used to recover the image and estimate its depth, which usually requires careful camera calibration and ad-hoc focus estimation \cite{grossmann_depth_1987}, acquisition of focal depth stacks (\cite{aguet_model-based_2008, shihavuddin_smooth_2017}), or coherent imaging, such as digital holographic microscopy \cite{cuche_simultaneous_1999}, to numerically refocus the image. Using an adequate \gls{psf}, i.e. one that corresponds to the blur, in a deconvolution algorithm can restore details in the image \cite{griffa_comparison_2010}. \gls{psf} estimation can be achieved by many techniques \cite{conchello_optical_2005}, but most of them are either dependent on a tedious calibration step, such as the experimental measurement of the PSF, or are sensitive to noise or image variability. \gls{bd} techniques are methods able to recover features in the image without prior knowledge of the \gls{psf}.

Here, we aim at estimating the local \gls{psf} only from the acquired image and use it to reverse the local degradation due to the optical system. Furthermore, we aim at estimating the depth of any location on the surface of a thin object with respect to the focal plane. We rely on a model-based approach that retrieves the \gls{psf} given a degraded image patch via a machine learning approach.
\\
Machine learning technologies have improved our ability to classify images \cite{krizhevsky_imagenet_2012}, detect objects \cite{ren_faster_2017}, describe content \cite{girshick_rich_2014}, and estimate image quality \cite{yang_assessing_2018}. \glspl{cnn}, in particular, have the ability to learn correlations between an image input and a defined outcome and appear well adapted to determining the degradation kernel directly from the image texture. A similar reasoning led to recent results by Zhu et al. \cite{zhu_estimating_2013}, Sun et al. \cite{sun_learning_2015}, Gong et al. \cite{gong_motion_2017} and Nah et al. \cite{nah_deep_2017}, where the direction and amplitude of motion blur was determined by a \gls{cnn} classifier from images blurred with a Gaussian kernel. Our approach is similar to that of \cite{sun_learning_2015} but with PSF models that are tailored to the specificities of microscopy, a concept that we initially introduced in \cite{shajkofci_semi-blind_2018} and that has since been used by other groups such as \cite{he_restoration_2020}. In particular, we considered a more generic physical model that can accommodate large-support \glspl{psf}. \glspl{cnn} were also used in a end-to-end manner to enhance details in biological images by performing supervised interpolation  \cite{weigert_content-aware_2018, rivenson_phase_2018, wu_three-dimensional_2019} or to emulate confocal stacks of sparse 3D structures from widefield images \cite{wang_deep_2019}.
\\
In this paper, we propose a method to:
\begin{enumerate}[nosep,  labelwidth=!, labelindent=0pt]
	\item Find the spatially-variant \gls{psf} of the degraded image of a thin, non-flat object directly from the image texture without any instrument-specific calibration step. The \gls{psf} determination technique is derived from the one we proposed in \cite{shajkofci_semi-blind_2018}, which recovers local Zernike moments of the \gls{psf}. We focus here on improving the degradation model and quantitatively assess the robustness of the method.
	\item Deconvolve the image in a blind and spatially-variant manner, using a regularized Richardson-Lucy algorithm with an overlap-add approach.
	\item Extract the depth of a three-dimensional surface from a single two-dimensional image using combinations of Zernike moments.
\end{enumerate}
This technique allows us to enhance the acquired image and recover the three-dimensional structure of a two-dimensional manifold in a 3D space using a single 2D image as an input.
  
This paper is organized as follows. In Section \ref{sec:methods}, we present the method, comprising the image formation model, the degradation model, the data set generation process, the different neural networks to be trained, the \gls{psf} mapping, the deconvolution algorithm, and the depth from focus algorithm. Then, in Section \ref{sec:experiments}, we characterize the regression performance of the \gls{cnn} for different modalities, as well as the gain in resolution from the deconvolution, and the precision of depth detection. We then discuss our findings in Section \ref{sec:discussion} and conclude in Section \ref{sec:conclusion}.
\section{Methods}
\label{sec:methods}
\subsection{Object and image formation model}
\label{sec:objectformation}
\begin{figure}
	\begin{minipage}{1.0\linewidth}
		\centering
		\centerline{\includegraphics[width=1.0\linewidth]{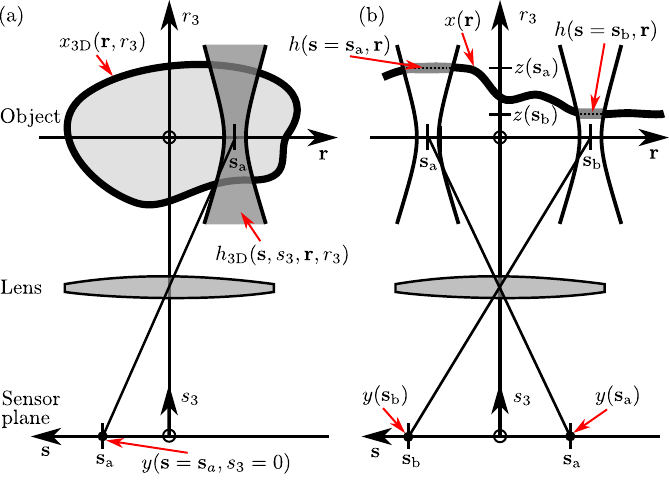}}
		\caption{Object and image formation model. (a) General thick 3D object case, (b) Approximate manifold case.}
		\label{fig:axis}
	\end{minipage}
\end{figure}

We consider a two-dimensional manifold in 3D-space with local intensity $x(\bm{r})$, e.g.\ an infinitely thin sample suspended in a gel, which can be parameterized by the lateral coordinates $\bm{r} = (r_1,r_2)$ and axial coordinate $z(\bm{r})$. We express the resulting three-dimensional object as:
\begin{equation}
	x_{3D}(\bm{r},r_3) = x(\bm{r}) \delta(r_3 - z(\bm{r})),
\end{equation}
where $(\bm{r},r_3) = (r_1,r_2,r_3)$ are coordinates in 3D object space.
We further consider an optical imaging system with camera coordinates $\bm{s} = (s_1, s_2)$ and axial position $s_3$, characterized by a spatially-varying point spread function $h_{\text{3D}}(\bm{s},s_3, \bm{r}, r_3)$ (see Fig.~\ref{fig:axis}). For a fixed axial camera position $s_3$, the measured intensity by a pixel at position $\bm{s}$ is given by the convolution (Fig.~\ref{fig:axis} (a)) \cite{goodman_chapter_2005}:
\begin{gather}
	\label{eq1}
	y_{\text{3D}}(\bm{s},s_3) = \iiint x_{\text{3D}}(\bm{r}, r_3) h_{\text{3D}}(\bm{s},s_3, \bm{r}, r_3) \text{d}\bm{r} \text{d}r_3,
\end{gather}
where we assumed, to simplify the notation, that the magnification is 1.
When the microscope is focused at the origin ($s_3=0$) we define the 2D image $y(\bm{s}) = y_{\text{3D}}(\bm{s},s_3) \bigr|_{s_3=0}$, which can be obtained via the expression:
\begin{align}
	y(\bm{s}) &= \iiint x(\bm{r}) \delta(r_3 - z(\bm{r})) h_{\text{3D}}(\bm{s},s_3 = 0, \bm{r}, r_3) \text{d}\bm{r} \text{d}r_3\\
	&= \iint x(\bm{r}) h_{\text{3D}}(\bm{s},s_3= 0, \bm{r}, r_3 = z(\bm{r})) \text{d}\bm{r}\\
	&= \iint x(\bm{r}) h(\bm{s},\bm{r}) \text{d}\bm{r},
\end{align}
where $h(\bm{s},\bm{r})=h_{\text{3D}}(\bm{s},s_3=0, \bm{r}, r_3 =z(\bm{r}))$ is a 2D point spread function that incorporates both the local (3D) variations of the optical system and the variable depth of the thin sample. We further assume that $h(\bm{s},\bm{r})$ can be approximated by a parametric function $\widetilde{h}_{\boldsymbol{a}(\bm{s})}(\bm{r})$, where the $N$ parameters $\boldsymbol{a}(\bm{s}) = \irow{a_1(\bm{s})&a_2(\bm{s})&\cdots &a_{N}(\bm{s})}$ can vary for every 2D location $\bm{s}$ of the image (Fig.~\ref{fig:axis} (b)).

\subsection{Parametric degradation models}
\label{sec:methods_psf_models}
There are many methods for estimating the \gls{psf} of a degraded image. Such methods can be categorized into two classes: direct \gls{psf} estimation or parametric modeling. In works such as those by Grossmann et al. \cite{grossmann_depth_1987}, the \glspl{psf} are estimated directly from the image (e.g. using edge detection and a regression model \cite{chalmond_psf_1991}, a \gls{map} prediction \cite{joshi_psf_2008}), or via a camera calibration using images of a defined and known pattern \cite{reuter_blurtags_2012} \cite{brauers_direct_2010}. Levin et al. \cite{levin_understanding_2011} showed that a \gls{map} approach to recover the blur kernel is well constrained, but that the \gls{map} global optimum for the recovered image is a blurred image because the strong constraints do not always generalize to unexpected or noisy types of data \cite{fish_blind_1995}, which are common in microscopy images. Full pixel-wise \glspl{psf} can also be estimated using dictionary learning \cite{soulez_learn_2014}, or \glspl{cnn} \cite{herbel_fast_2018}. However, this latter kind of estimation is not well constrained and can generate over-fitting artifacts.

In contrast, parametric modeling of the \gls{psf} allows to reduce the dimensionality of the optimization problem and to attach a physical meaning to the parameters, such as the relative distance from the focal point, or optical aberrations such as astigmatism. There are many mathematical models to represent the \gls{psf} of a microscope. They can take into account both physical characteristics of the objectives (for example numerical aperture, correction types, etc.) and of the experimental conditions (focal distance and immersion medium) \cite{aguet_accurate_2008}. In many cases, the model parameters correspond to physical design conditions, such as optical distances, aperture diameters, or foci. A simple \gls{psf} model can be obtained from the Fraunhofer diffraction theory to calculate the diffraction of a circular aperture \cite{airy_diffraction_1835}. The \emph{Gibson \& Lanni} model accounts for the immersion medium, the cover-slip, the sample layers, different medium numerical apertures, and the properties of the objective \cite{gibson_diffraction_1989}. Despite their theoretical relevance, in practice, numerical values for these parameters may not be available, as detailed information about all experimental and design conditions may be lacking. Even if the parameters are accessible, they may be missing if they were not recorded along with the image.

Since we aim to recover the \gls{psf} from the image itself, with minimal knowledge of the imaging conditions, we focused on models specified by only a small number of parameters or whose complexity can be adjusted progressively by considering approximations with a subset of the complete set of parameters. Specifically, we considered \gls{psf} models based on Zernike polynomials, which are used to describe the wavefront function of lenses such as the eye \cite{iskander_optimal_2001}, as well as anisotropic Gaussian models.
\subsubsection{Zernike polynomial decomposition of the pupil function}
\label{sec:zernike}
Optical abnormalities, such as de-focus, astigmatism, or spherical aberrations, can be modeled with a superposition of Zernike polynomials $Z_n(\bm{\xi})$ in the expansion of the microscope objective's pupil function $W_{\boldsymbol{a}(\bm{s})}$ \cite{von_zernike_beugungstheorie_1934}:
\begin{equation} 
	W_{\boldsymbol{a}(\bm{s})}(\bm{\xi}) = \sum_{n=1}^{N}  Z_n(\bm{\xi}, a_n(\bm{s})), 
\end{equation}
where $\bm{\xi}$ denotes the two-vector of spatial coordinates in the pupil plane perpendicular to the optical axis, $N$ the maximal order of considered aberrations, and $a_n(\bm{s})$ the parameter corresponding to each Zernike term $Z_n$. In our experiments, $Z_1$ describes the de-focus term, $Z_2$ describes the power of the astigmatism (cylinder), and $Z_3$ describes the astigmatism angle (axis).
The pupil function can ultimately be converted into a \gls{psf} \cite{goodman_introduction_2005}:
\begin{equation} \label{eq:pdffromzernike}
	\widetilde{h}_{\boldsymbol{a}(\bm{s})}(\bm{r}) \propto \left| \mathcal{F}\left\{W_{\boldsymbol{a}(\bm{s})}\left(\frac{\bm{r}}{\lambda} \right)\right\} \right|^2,
\end{equation}
with $\mathcal{F}$ the Fourier transform and $\lambda$ the wavelength of the light.
\subsubsection{Anisotropic Gaussian model}
\label{sec:gaussian}
For many applications, Gaussian distributions are sufficiently accurate approximations of the diffraction-limited \gls{psf} of wide field microscopes \cite{zhang_gaussian_2007}. We extend the model by allowing for anisotropy, which we require to describe astigmatic aberrations, or if the spatial resolution in one lateral direction is different from that in the other. The detection method is in that case similar to the one described in \cite{sun_learning_2015}.
The \gls{psf} is then defined by the anisotropic zero-centered normal probability density function:
\begin{multline}
	\widetilde{h}_{\boldsymbol{a}(\bm{s})}(\bm{r}) = \frac{1}{\sqrt{2\pi a_1(\bm{s})}} \exp\left({-\frac{1}{2} \frac{{r_1}^2}{a_1(\bm{s})}}\right)
	\\
	\cdot \frac{1}{\sqrt{2\pi a_2(\bm{s})}} \exp\left({-\frac{1}{2} \frac{{r_2}^2}{a_2(\bm{s})}}\right),
\end{multline}
where $a_1$ and $a_2$ are the variances of the Gaussian in the x and y axes, respectively.
\subsection{Problem statement}
We aim to solve several problems. First, given only an observed degraded image \textbf{$y(\bm{s})$}, we want to estimate the \gls{psf} model $\widetilde{h}_{\boldsymbol{a}(\bm{s})}(\bm{r})$ closest to the effective \gls{psf} of the imaging system $h(\bm{s},\bm{r})$ for any point $\bm{s}$ \emph{without} requiring additional information on the microscope or any further calibration images acquired with that microscope. Specifically, we want to infer the model parameters $\boldsymbol{a}(\bm{s})$, first locally, then globally. Next, given the local \gls{psf} parameters $\boldsymbol{a}(\bm{s})$ and the blurred image $y(\bm{s})$, we want to recover an estimate of the non-degraded image \textbf{$x(\bm{r})$}. Finally, we want to infer the local depth $z(\bm{r})$ along the axis of the object at any position $\bm{r}$ in the plane perpendicular to the optical axis 
thereby allowing us to build $x_{\text{3D}}(\bm{r}, z(\bm{r}))$. 

\subsection{Method overview}
For each of the problems, we summarize the following main steps:
\begin{enumerate}
    \item Shift-invariant \gls{psf} parameter estimation given an image patch (see Section \ref{sec:psf_estimation_methods_loss})
    \begin{enumerate}
		\item Select a parametric degradation model for $h_{\boldsymbol{a}}(\bm{r})$ allowing the generation of \gls{psf}/parameters pairs. 
		\item Gather a training library of microscopy images, degrade each image via a spatially-\emph{invariant} convolution with its corresponding \gls{psf}, corrupt it with synthetic noise.
		\item Train a \gls{cnn} that takes a degraded image patch as input and returns the corresponding degradation model parameters, via regression.
	\end{enumerate}
	\item Local \gls{psf} estimation given a full degraded microscopy image (see Section \ref{sec:methods_psfmapping})
	\begin{enumerate}
	 	\item Given a full microscopy image as input, locally extract a patch, then regress the \gls{psf} parameters using the steps above.
		\item Repeat in all regions of the image.
		\item Combine the estimated \gls{psf} parameters to generate the map $\boldsymbol{a}(\bm{s})$ of the local \gls{psf} model parameters.
	\end{enumerate}
	\item Application 1: spatially-variant blind deconvolution (see Section \ref{sec:application1})
	\begin{enumerate}
		\item Given an input image and the map of estimated local \gls{psf} parameters, generate local \glspl{psf} $\widetilde{h}_{\boldsymbol{a}(\bm{s})}(\bm{r})$.
		\item Use the generated \glspl{psf} in a \gls{tvrl} deconvolution algorithm to recover an estimate for $x(\bm{r})$.
	\end{enumerate}
	\item Application 2: estimate depth from focus using \gls{psf} engineering (see Section~\ref{sec:application2}))
	\begin{enumerate}
		\item Given an image acquired with a \gls{psf}-engineered optical system $y(\bm{s})$ \cite{kao_tracking_1994}), deduce the depth map $z(\bm{r})$ from the local parameters $\boldsymbol{a}(\bm{s})$. 
	 	\item Generate $x_{\text{3D}}(\bm{r}, z(\bm{r}))$.
	 \end{enumerate}
\end{enumerate}
In the following subsections, we provide details on each of these steps.

\subsection{\gls{psf} parameter estimation in image patches (shift-invariant image formation model)}
\subsubsection{Data set generation for \gls{cnn} training}
\label{sec:methods_dataset_generation}
\begin{figure*}
	\begin{minipage}{1.0\linewidth}
		\centering
		\centerline{\includegraphics[width=\linewidth]{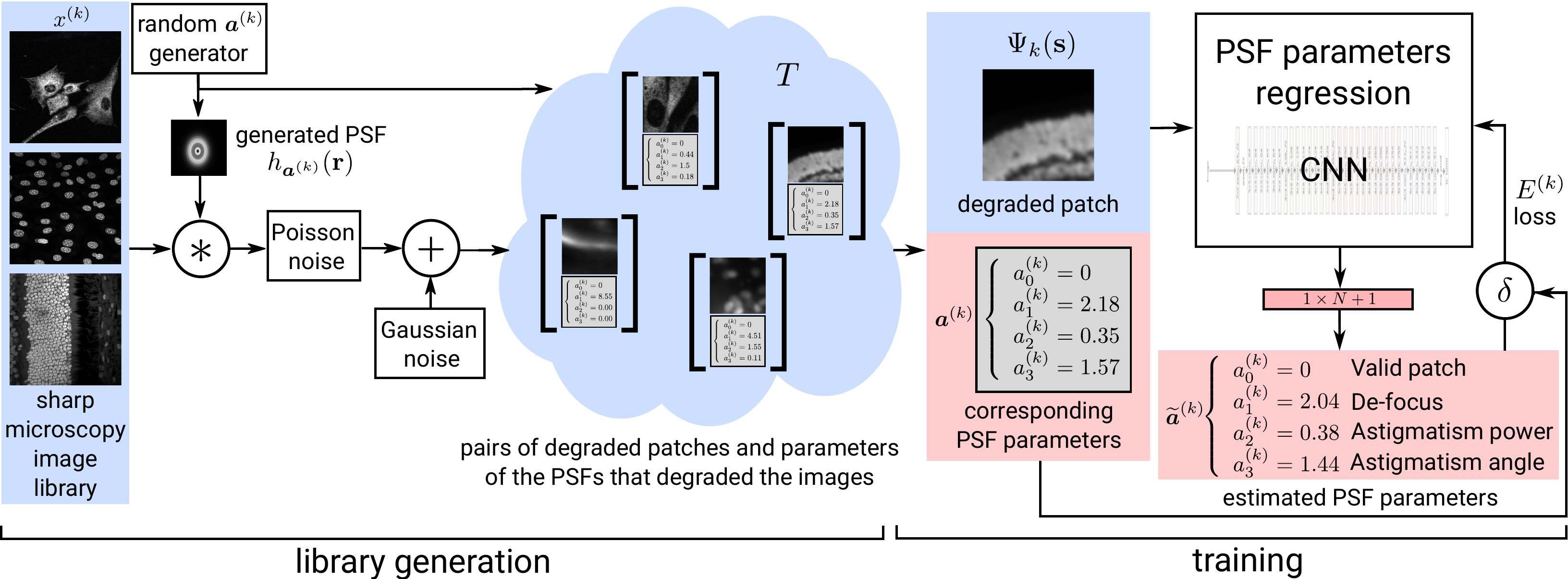}}
		\caption{Data set gathering (left) and \gls{cnn} training (right) pipelines. From a large library of sharp microscopy images, small patches are created, blurred with a \gls{psf} generated from random parameters, and degraded with a Poisson-Gaussian noise mixture (see Section \ref{sec:methods_dataset_generation}). The resulting patches and the parameters are stored in the training set $T$, that is used for training the \gls{cnn}. Using backpropagation of the loss function, the \gls{cnn} output is trained towards the prediction of the \gls{psf} model parameters (Section \ref{sec:psf_estimation_methods_loss}).}
		\label{fig:methods_training}
	\end{minipage}
\end{figure*}

Given an image patch as input, we wish to estimate the degradation model parameters corresponding to the spatially-invariant \gls{psf} that degraded the patch. Since neural networks are trained by adjusting their internal weights using backpropagation of the derivative of a loss function between the ground truth of the training data and the output of the network \cite{lecun_efficient_2012}, we establish a training set $T$ (Fig. \ref{fig:methods_training}). For that purpose, we gather input images $x^{(k)}$. Since reliable public image data sets have limited size, we augment the size of the training sets by rotating each image by an angle of $\pm 90^{\circ}$, so we have $K$ images, $x^{(k)}$, $k = 0, ... , K-1$.. We induce a synthetic degradation by convolving the images with $K$ generated \glspl{psf} $h_{\boldsymbol{a}^{(k)}}$ of model parameters $\boldsymbol{a}^{(k)}$ drawn from a normalized and scaled uniform distribution:
\begin{equation}
	\psi^{(k)}(\bm{s}) = (h_{\boldsymbol{a}^{(k)}} \ast x^{(k)})(\bm{s}).
\end{equation}
We also consider the two predominant sources of noise in digital image acquisition: the stochastic nature of the quantum effects of the photoconversion process and the intrinsic thermal and electronic fluctuations in the CCD camera \cite{luisier_image_2011}. The first source of noise comes from physical constraints such as a low-power light source or short exposure time, while the second is signal-independent. This motivates the noise model as a mixed Gaussian-Poisson noise process.
Therefore, we define noise with the two following components:
\begin{itemize}
	\item A random variable $n_p(\bm{s}) \sim \mathcal{P}(\lambda = \psi^{(k)}(\bm{s}))$ following a Poisson distribution of probability $P\{n_p(\bm{s}) = i\} = e^{-\lambda} \lambda^i / {i!}$.
	\item A random variable $b(\bm{s}) \sim \mathcal{N}(0, \sigma^2)$ following a Gaussian distribution with zero-mean and variance $\sigma^2$.
\end{itemize} 
The image noise model for data set generation is then:
\begin{equation}
	\label{eq:methods_degradation}
	\psi_{\text{noisy}}^{(k)}(\bm{s}) = \beta n_p(\bm{s}) + b(\bm{s}),
\end{equation} 
with $\beta$ a number between 0 and 1 reflecting the quantum efficiency of the CCD \cite{gilbert_amplitude_1960}. Images that did not comply with a minimal variance and white pixel ratio were tagged as invalid, i.e. $a^{(k)}_{0} = 1$ (see Section \ref{sec:psf_estimation_methods_loss}).

We cropped the images $\psi_{\text{noisy}}^{(k)}(\bm{s})$ to a size  $K_\psi \times L_\psi$ by randomly selecting the position of a region of interest of that size. We then paired these image patches with their respective \gls{psf} parameter vector $\boldsymbol{a}^{(k)}$ in order to form the training set $T = \{(\psi_{\text{noisy}}^{(k)}(\bm{s}), \boldsymbol{a}^{(k)})\}_{k=0}^{K-1}$.

\vspace{6pt}
\subsubsection{\gls{cnn} training modalities}
\label{sec:psf_estimation_methods_loss}
We considered several neural networks (whose architectures we further describe below) and trained them to learn the \gls{psf} model parameters described in Section \ref{sec:methods_psf_models}. The task of the network is to estimate, only from the $k^{\text{th}}$ input image patch $\Psi^{(k)}(\bm{s})$, the parameters $\widetilde{\bm{a}}^{(k)}(\bm{s})$ that have been used by the \gls{psf} model to degrade that input image. Since there are cases where the \gls{psf} estimation is not possible, e.g. where the sample lacks texture, such as in uniformly black or grey areas, we added a boolean parameter $a^{(k)}_{0}$ (whose values can be either $0$ or $1$), which indicates the legitimacy of the sample. The total number of estimated parameters is then $N+1$. We aim at minimizing the distance between the output of the network $\widetilde{\bm{a}}^{(k)}$ and the ground-truth \gls{psf} parameters ${\bm{a}^{(k)}}$. Therefore, in the training phase, we updated the weights of the \gls{cnn} using the modified Euclidean loss function:
\begin{gather}
	E^{(k)} = \gamma \left({a^{(k)}_{0}} - {\widetilde{a}^{(k)}_{0}}\right)^2 + \frac {1-{a^{(k)}_{0}}} {2N} \sum_{n=1}^{N}( {a^{(k)}_{n}} - {\widetilde{a}^{(k)}_{n}} )^2 ,
\end{gather}
with $\gamma$ a hyperparameter regulating the importance of the validity parameter, that we set to $1$ in our further experiments.
\ignore{where ${\widetilde{a}^{(k)}_{n}}$ is the estimate of the $n$-th parameter of the \gls{psf} model applied on the training image $\Psi^{(k)}$, and ${a^{(k)}_{n}}$ the ground-truth parameter as stored in $T$.}

We choose to rely on networks that showed good performance in the ImageNet competition, which is a benchmark in object classification on hundreds of categories \cite{deng_imagenet_2009,russakovsky_imagenet_2015}. Donahue et al. \cite{donahue_decaf_2014} showed that deep convolutional representations can be applied to a variety of tasks and detection of visual features, which drove our selection for estimating optical aberrations. Hendrycks et al. \cite{hendrycks_benchmarking_2019} extensively discussed whether the networks were robust to changes in input illumination, noise, and blur. While residual networks that use skip-connections such as ResNet \cite{he_deep_2016} appear to be more robust to input noise than primitive feedforward networks such as AlexNet \cite{krizhevsky_imagenet_2012}, their performance appears to be surpassed by newer multibranch models such as ResNeXt \cite{xie_aggregated_2017} or Densenet \cite{huang_densely_2017}. In the context of our specific task, we compared the performance (see Section \ref{sec:results_accuracy}) and the robustness to degradation (see Section \ref{sec:results_robustness}) of several of the above architectures.

After training, the networks can regress the spatially-invariant \gls{psf} parameters $\widetilde{\bm{a}}^{(k)}(\bm{s})$ from a single input image patch $\Psi^{(k)}(\bm{s})$.

\subsection{Spatially-variant \gls{psf} parameter mapping}
\label{sec:methods_psfmapping}
\begin{figure}
	\begin{minipage}{1.0\linewidth}
		\centering
		\includegraphics[width=\linewidth]{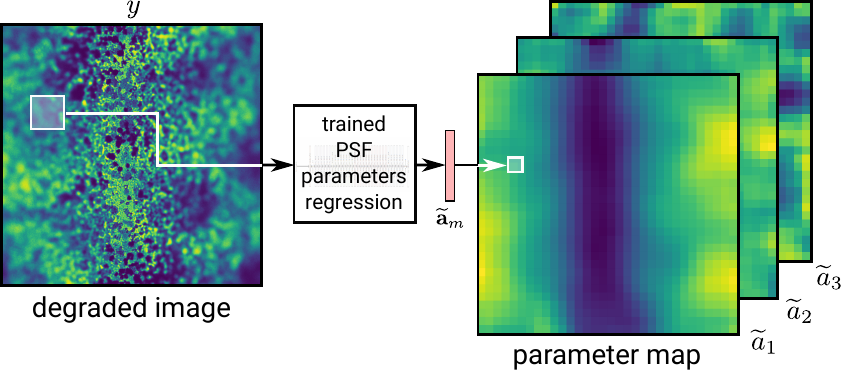}
		\caption{Spatially-variant \gls{psf} parameter mapping using a sliding window over the acquired image, as input of a convolutional neural network.}
		\label{fig:psfmap}
	\end{minipage}
\end{figure}

Given a trained \gls{cnn} that is able to recover the degradation parameters from a single image patch, we now turn to the problem of locally estimating the parameters of the different \glspl{psf} that degraded a larger input image. To achieve this, we use an overlapping sliding window over the input image $x(\bm{r})$ with stride $t$ that is fed into the locally invariant regression \gls{cnn} trained in Section \ref{sec:psf_estimation_methods_loss} (see Fig. \ref{fig:psfmap}). We store the resulting parameters in the map $\widetilde{\bm{A}} = \irow{\widetilde{\bm{a}}^{(0)} & \widetilde{\bm{a}}^{(1)} & ... & \widetilde{\bm{a}}^{(M)}}$, where $\widetilde{\bm{a}}^{(m)}$ is the output of the neural network for patch $m$ and $M$ is the total number of patches. Using the \gls{psf} model, we generate from $\widetilde{\bm{A}}$ a spatially-variant map of local \gls{psf} kernels defined as $\widetilde{{H}} = \irow{\widetilde{{h}}^{(0)}(r) & \widetilde{{h}}^{(1)}(r) & \cdots & \widetilde{{h}}^{(M)}(r)}$.
\\
The overlapping window over the input image yields a map of $\left( \lfloor ({K_x - K_\psi})/t\rfloor+1 \right) \times \left( \lfloor ({L_x - L_\psi})/t\rfloor+1 \right)$ kernels, with $K_x$, $L_x$, and $K_\psi$, $L_\psi$ being the width and height of the input image and the window step size, respectively. For example, a $1024\times 1024 $ pixel input image using $128 \times 128 $ pixel patches and $t=64$ yields $M = 13 \times 13$ spatially-dependent \gls{psf} kernels. We fill every patch with a validity parameter $a_{0} = 1$ (i.e. invalid) with the content of the inverse Euclidean distance-weighted average of the four-connected nearest neighbors using K-Nearest Neighbor regression \cite{hechenbichler_weighted_2004} in order to avoid boxing artifacts during the later deconvolution process.
\begin{figure}
	\begin{minipage}{1.0\linewidth}
		\centering
		\centerline{\includegraphics[width=\linewidth]{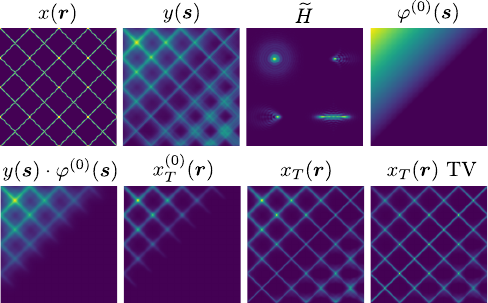}}
		
		\caption{Synthetic experiment involving variables and results of Eq. (\ref{eq:methods_deconvolution}). Starting from a ground truth image ($x(\bm{r})$), a local map of \gls{psf} (smooth interpolation between the 4 shown \glspl{psf}) and local weight combination $\bm{\varphi}(\bm{s})$, we generated a blurred image y. The deconvolution method (similar to \cite{hirsch_efficient_2010} and \cite{dey_richardsonlucy_2006}) starts from a map of locally estimated \glspl{psf} to deconvolve and recombine into a single image $x_T(\bm{r})$ and $x_T(\bm{r})$ TV.}
		\label{fig:deconvolution_synthetic}
	\end{minipage}
\end{figure}
\subsection{Application 1: spatially-variant blind deconvolution}
\label{sec:application1}
We now turn to the problem of deconvolving the image.
Existing deconvolution techniques can be categorized into three classes: (1) Non-blind methods, (2) entirely blind methods, and (3) parametric semi-blind algorithms. Non-blind methods require full knowledge of the \gls{psf} (\cite{soulez_learn_2014}, \cite{weigert_isotropic_2017}), while the latter two classes aim at improving the image without prior knowledge of the \gls{psf}, the object, or other optical parameters. Entirely blind algorithms, such as \cite{hirsch_efficient_2010} are based on optimization and estimation of the latent image or kernel \cite{levin_understanding_2011}. Many blind deconvolution techniques are computationally expensive, especially for larger convolution kernels, and assume spatially invariant \glspl{psf}.

Parametric or semi-blind algorithms are blind methods that are constrained by knowledge about the transfer function distribution, such as a diffraction model or a prior on the shape of the \gls{psf} (\cite{aguet_model-based_2008}, \cite{morin_semi-blind_2013}). Parametric models allow reducing the complexity of the optimization problem, increasing the overall robustness, and avoiding issues such as over-fitting. However, it remains difficult to estimate the parameters from experimental data. We will focus on this third class of deconvolution methods, since our local \gls{psf} estimation method allows inferring their parameters without measuring any of them experimentally. Since we set a diffraction model defining the shape of the local \gls{psf} in a large image, we describe our deconvolution method as semi-blind.

Given the degraded image and a local map of \gls{psf} parameters, we restore the input using \gls{tvrl} deconvolution. \gls{rl} is an iterative maximum-likelihood approach and assumes that the noise follows a Poisson distribution \cite{richardson_bayesian-based_1972}, which is well adapted for microscopy. The method is subject to noise amplification, which can, however, be counterbalanced by a regularization term that penalizes the $l_1$ norm of the gradient of the signal (\cite{chan_total_1998}, \cite{dey_richardsonlucy_2006}). Here, we assume that the \gls{psf} is spatially invariant in small parts of the image. Spatially-variant convolution techniques have been extensively reviewed by Denis et al. \cite{denis_fast_2015}. Hirsch et al. \cite{hirsch_efficient_2010} have shown that the local invariance assumption can be improved by filtering the input with every local \gls{psf} and then reconstructing the image using interpolation.
We extend this method by its inclusion in the \gls{tvrl} algorithm. Rather than interpolating deconvolved images, the overlap-add filtering method, as described in \cite{hirsch_efficient_2010, temerinac_ott_spatially_2011}, interpolates the \gls{psf} for each point in the image space.
The idea for such method is: (i) to cover the image with overlapping patches using smooth interpolation, (ii) to deconvolve each patch with a different \gls{psf}, (iii) to add the patches to obtain a single large image. The equivalent for convolution can be written as:
\begin{equation}
x(\bm{r}) = \sum_{m = 0}^M (\widetilde{h}^{(m)} * (\varphi^{(m)} \odot y))(\bm{r}),
\label{eq:methods_interpolation_psf}
\end{equation}
where $\varphi^{(m)}(\bm{s})$ is the masking operator of the $m$th patch. We illustrated the masking and deconvolution steps in Fig. \ref{fig:deconvolution_synthetic}.
Since the RL algorithm tends to exacerbate edges and small variations such as noise, we use \gls{tv} regularization to obtain a smooth solution while preserving the borders \cite{dey_richardsonlucy_2006}. The image at each \gls{rl} iteration becomes:
\begin{multline}
	x_{i+1}(\bm{r}) = \sum_{m = 0}^M
	\left[\frac{(\widetilde{h}^{(m)} * (\varphi^{(m)} \cdot y))(\mathord{\bm{r}})}{(\widetilde{h}^{(m)} * x^{(m)}_i)(\mathord{\bm{r}})}
	* {\widetilde{h}^{(m)}}(\mathord{-\cdot})\right]
	\\
	\cdot
	\frac{x^{(m)}_i(\bm{r})}
	{1 - \lambda_{TV} \hspace{2pt}\text{div}\left(\frac{ \triangledown x^{(m)}_i(\bm{r})}{\lvert\triangledown x^{(m)}_i(\bm{r}) \rvert}\right)},
	\label{eq:methods_deconvolution}
\end{multline}
with $y(\bm{s})$ the blurry image, $x_i(\bm{r})$ the deconvolved image at iteration $i=1,...,I$, $x^{(m)}_i(\bm{r})$ the $K_y \times L_y$ deconvolved patch at iteration $i$, $M$ the number of patches (and different \glspl{psf}) in one image $x$, $\widetilde{h}^{(m)}$ the $K_h \times L_h$ \gls{psf} for patch $m$ and $\lambda_{TV}$ the \gls{tv} regularization factor. $\triangledown x_i^j(r)$ is the finite difference operator, which approximates the spatial gradient.\ignore{ If $\lambda_{TV}$ is small, \gls{rl} is dominated by the data, and if $\lambda_{TV} \sim 1$, \gls{rl} is dominated by the regularization \cite{dey_richardsonlucy_2006}. }
\subsection{Application 2: depth from focus using astigmatism}
\label{sec:application2}
The spatially-variant \gls{psf} parameter mappings obtained in Section \ref{sec:methods_psfmapping} yield local parameters, such as the blur, that are a function of the distance of the object to the focal plane. However, due to the symmetry of the \gls{psf} in depth, this function is ambiguous about the sign of the distance map $z(\bm{s})$ (above or below the focal plane). That is why we now aim at estimating the depth map $z(\bm{s})$ for every lateral pixel $\bm{s}$ of the 2D manifold in 3D space using our trained neural network and one single image as input. To achieve this, we use a local combination of Zernike polynomial coefficients $\boldsymbol{a}(\bm{s})$. The de-focus coefficient $a_1(\bm{s})$ is linked to the distance of the object to the focal plane, but there is no information about whether the object is in the front or behind the focal point. To address this limitation, we took inspiration from several methods to retrieve the relative position of a particle by encoding it in the shape of its \gls{psf} (either via use of astigmatic lenses (\cite{kao_tracking_1994, huang_three-dimensional_2008}) or by use of a deformable mirror to generate more precise and complex \gls{psf} shapes \cite{aristov_zola-3d_2018}).

\begin{figure}
	\begin{minipage}{1.0\linewidth}
		\centering
		\centerline{\includegraphics[width=1.0\linewidth]{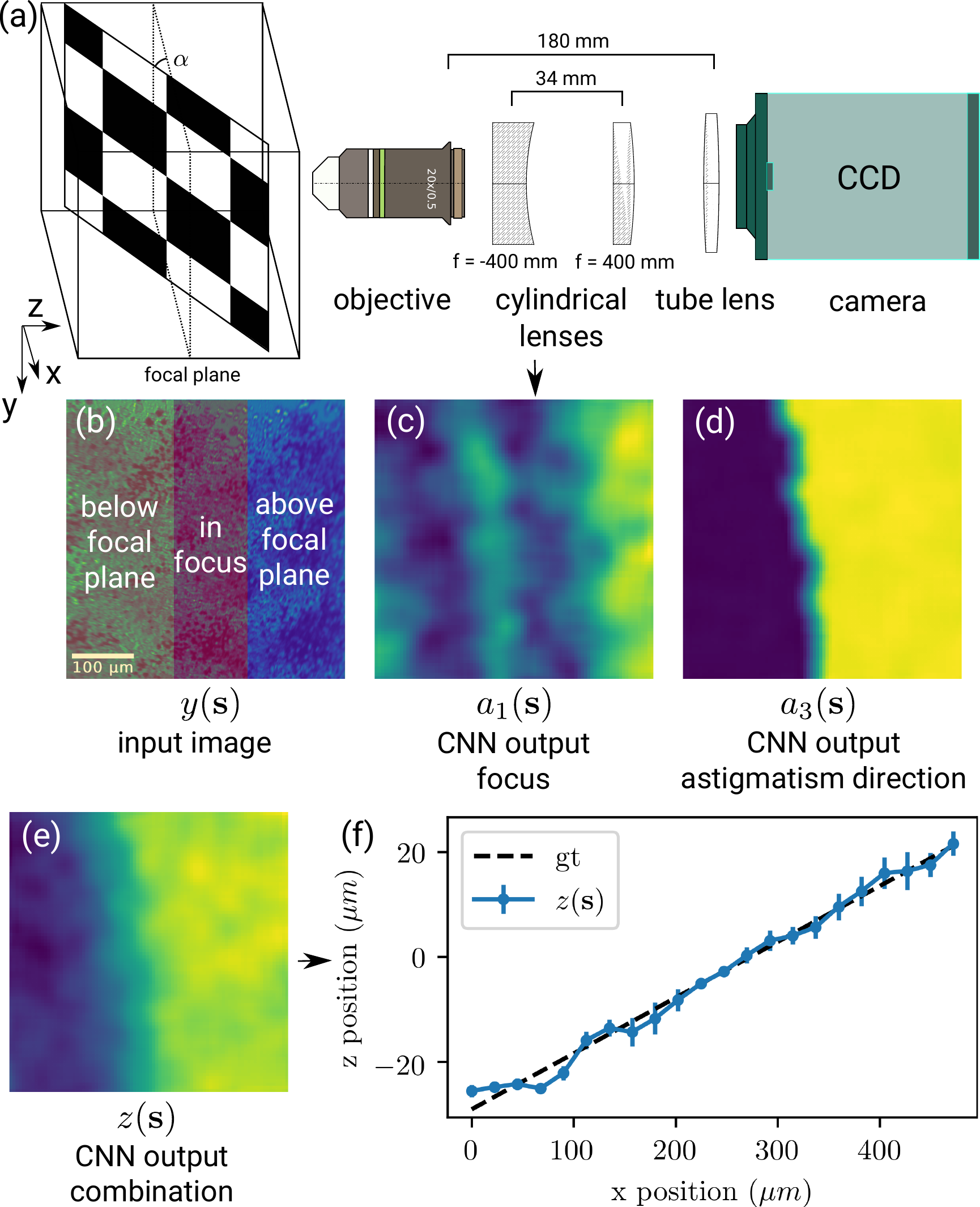}}
		\caption{Depth estimation of a plane using controlled astigmatic aberrations by use of cylindrical lenses. (a) Optical system with using cylindrical lenses in the infinite tube space to induce astigmatism. (b) Image of a grid taken from the camera, with highlighted parts of the surface above  and below the focal plane. (c),(d) Output of the \gls{cnn} using (b) as input. The resolution of the map is $M=31\times 31$ different \glspl{psf}. (e) Weighted combination of (c) and (d) to form the depth map. (f) Projection of the depth of the surface over the $y$ axis and comparison to the ground truth data.}
		\label{fig:application2}
	\end{minipage}
\end{figure}

We used two cylindrical lenses of focal length $\SI{-400}{\micro\metre}$ and $\SI{400}{\micro\metre}$, separated by $\SI{3.4}{\centi\metre}$ thereby giving a combined focal length of $f = \SI{6000}{\milli\meter}$ and placed them in the infinity space of a microscope to generate an imaging system with an astigmatic \gls{psf} (Fig. \ref{fig:application2}). We used the networks trained in Section \ref{sec:methods_dataset_generation} using 2D-Zernike models to infer the depth map $z(\bm{s})$ from the 2D image $y(\bm{s}$) of the tridimensional surface $x_{\text{3D}}(\bm{r}, z(\bm{r}))$. We defined a distance metric by multiplying the output focus parameter and the normalized and zero-centered astigmatism direction:
\begin{equation}
\label{eq:application2}
z(\bm{s}) = a_1(\bm{s}) \left(\frac{2 a_3(\bm{s})}{\pi} -1\right),
\end{equation}
with $a_1(\bm{s})$ the spatially local de-focus Zernike coefficient, and $a_3(\bm{s}) \in (0, \pi)$ the spatially local Zernike coefficient encoding the direction of astigmatism.

\section{Experiments}
\label{sec:experiments}
In this section we report our efforts to characterize the performance of our method as well as its dependency to several hyper-parameters, such as the choice of the \gls{psf} model, the neural network architecture, the training set size, or its content. We furthermore tested the regression performance of the \gls{psf} parameter regression and its robustness to \gls{snr} degradation and the absence of texturing. Finally, we also assessed the quality of deconvolved images and the accuracy of the estimated depths. 

\subsection{Infrastructure}
Our \gls{psf} parameter estimation method depends on three main variables: the content and size of training data sets, the \gls{psf} parametric model and the neural network architecture. We briefly describe the different options below.
\subsubsection{Training, validation and test data sets for \gls{cnn} regression performance}
\begin{figure}
	\begin{minipage}{1.0\linewidth}
		\centering
		\centerline{\includegraphics[width=\linewidth]{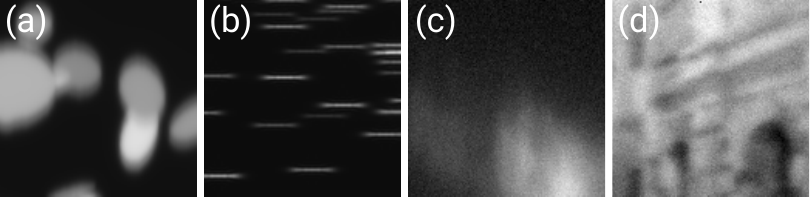}}
		\caption{Examples of degraded input patches from different data sources: (a) synthetic cells [syn], (b) synthetic points [poi], (c) microscopy images [micr], (d) natural images [nat]. The images have been degraded by the Zernike-3 \gls{psf} model (see Table \ref{fig:psfmodels}) and noise as described in Section \ref{sec:methods_dataset_generation}}.
		\label{figsynthetic}
	\end{minipage}
\end{figure}

We gathered images from four different data sources:
\begin{enumerate}
	\item \textbf{[micr]} microscopy images collected from \cite{bray_workflow_2012}, \cite{ljosa_annotated_2012} and \cite{williams_image_2017},
	\item \textbf{[nat]} common images from the MIT Places365 data set \cite{zhou_places_2018} that gathers natural and man-made photographs,
	\item \textbf{[poi]} synthetic images of points on a black background (Fig. \ref{figsynthetic}),
	\item \textbf{[syn]} synthetic images of cells (Fig. \ref{figsynthetic}).
\end{enumerate}
The rationale for using natural and synthetic images is that these data sources are much more abundant than microscopy images, often sharper and royalty free, making it possible to quickly assemble a large dataset. We combined these data to generate six different data sets (Table \ref{figdataaug}) and prepared the library as described in Section \ref{sec:methods_dataset_generation}.\ignore{These patches were further degraded by different \gls{psf} models. The training data sets are made of pairs of images patches and the parameters of the source image degradation (respectively $\Psi^{(k)}$ and ${\bm{a}^{(k)}}$).} We randomly selected two times 10,000 images to form a validation set and an test dataset that the networks never use during the training process. The validation dataset is used for selecting the best learning rate and early stopping epoch for every training, while the testing set is used for performance assessment. We added synthetic black images to every data set to avoid misdetection of non-textured parts of the image and explicitly set  ${a^{(k)}_{0} = 1}$ for these samples.

\begin{table}
	\caption{Name and size of the different training, validation and test data sets as input of the \gls{cnn}.}
	\begin{minipage}{1.0\linewidth}
		\centering
								\scalebox{0.9}{
			\begin{tabularx}{1.12\linewidth}{XXXX} \toprule
				Data set & $K_{train}$ & $K_{valid}$ & $K_{test}$\\\midrule
				syn  & 440,000 & 10,000 & 10,000\\
				poi  & 330,000& 10,000 & 10,000\\
				micr & 2,700,000 & 10,000 & 10,000\\
				micrsm & 270,000 & 10,000 & 10,000\\
				nat & 2,700,000 & 10,000 & 10,000\\
				micr-syn-poi & 3,470,000 & 10,000 & 10,000\\
				\bottomrule
			\end{tabularx}
}
		\label{figdataaug}
	\end{minipage}
\end{table}

\subsubsection{\gls{psf} models and parameters}
We considered two different \gls{psf} model types: Zernike polynomials (Section \ref{sec:zernike}) with $N=1,2,$ or $3$ parameters, and Gaussian (Section \ref{sec:gaussian}), with either $N=1$ or 2 parameters, as described in Table \ref{fig:psfmodels}.
\begin{table}
	\begin{minipage}{1.0\linewidth}
		\centering
				\caption{\gls{psf} models selected for data set generation (Section \ref{sec:methods_dataset_generation}), with the number and name of free parameters.}
						\scalebox{0.8}{
			\begin{tabularx}{1.26\linewidth}{ccY} \toprule
				\gls{psf} model & $N$ & Parameters\\\midrule
				Zernike-1 (Z-1)& 1 & focus\\
				Zernike-2 (Z-2)& 2 & cylinder, axis\\
				Zernike-3 (Z-3)& 3 & focus, cylinder, axis\\
				Gaussian-1 (G-1)& 1 & width\\
				Gaussian-2 (G-2)& 2 & width x axis, width y axis\\
				\bottomrule
				
			\end{tabularx}
		}

		\label{fig:psfmodels}
	\end{minipage}
\end{table}

\subsubsection{\gls{cnn} architectures and training modalities}
We compared two residual neural networks architectures trained from scratch: 34-layer ResNet \cite{he_deep_2016} and 50-layer ResNeXt \cite{xie_aggregated_2017}. They were adapted for accepting normalized gray-scale input image patches of size $W_\psi \times H_\psi = 128 \times 128 $ pixels. Additionally, we fine-tuned, using our training dataset, the same network already trained on the ImageNet data set (available on the PyTorch website). For the latter model, we re-scaled the input images to the network input size using bilinear interpolation. We trained the models for 20 epochs with PyTorch 1.0 using the Adam optimizer \cite{kingma_adam_2015} and a learning rate between $0.001$ and $0.01$ defined by the validation set performance.

%
%
%
		\label{fig:cnnmodels}

\subsection{Characterization of the \gls{cnn} regression performance}
\label{sec:results_accuracy}
We analyzed the performance of our system for regressing the \gls{psf} parameters. The metrics we used to assess the performance of the network is the goodness-of-fit of the parameter estimation compared to the ground truth. We quantified it in terms of the squared Pearson correlation coefficient $R^2$ averaged over all \gls{psf} parameters:
\begin{equation}
R^2 = \frac{1}{N}\sum_{n=0}^{N}{\left(1 - \frac{\sum_{k = 0}^{K_{\text{test}}-1}{(a^{(k)}_{n} - \widetilde{a}^{(k)}_{n})^2}}{\sum_{k = 0}^{K_{\text{test}}-1}{(a^{(k)}_{n} - \bar{a}_{n})^2}}\right)},
\end{equation} with $K_{\text{test}}$ the number of samples in the test set.
We calculated the correlation coefficient only for samples that contained texture in the ground-truth (i.e. when $a^{(k)}_{0} = 0$) and discarded the others.
\\
\subsubsection{Characterization of \gls{cnn} regression performance when training and test data set types are the same}
We started by assessing the performance of the \glspl{cnn} when the test set is made of the same image type as the training set. Table \ref{figcnnresults} summarizes the performance of the regression of test data for every combination of training data sets (Table \ref{figdataaug}), \gls{psf} models (Table \ref{fig:psfmodels}) and \gls{cnn} architectures.

\begin{framed}
	\textbf{Variables:} data set type, \gls{psf} model type, network architectures.
	\\
	\textbf{Fixed:} the data set type is the same for training and testing.
	\\
	\textbf{Evaluation criterion:} $R^2$ between the degradation parameters used to generate the test image and the parameters recovered by the \gls{cnn}.
\end{framed}

In most cases, the correlation coefficient is superior to 80\%, which indicates a very good degree of overall correlation. The worst cases are with models trained for Zernike-3, that yield $0.61 < R^2 < 0.96$. We notice a few differences in the regression performances between Gaussian and Zernike models. Indeed, images blurred with a Gaussian model tend to be better recognized by the neural network, with $R^2 > 0.90$, than images blurred with a Zernike model that fluctuates around $0.60 < R^2 < 1.00$. When looking at the performance of a smaller [micr] training data set compared to the full [micr] data set, we notice that the performance of the smaller data set is always worse or equal, no matter which \gls{cnn} model or \gls{psf} model used. Finally, we observe that the overall performance of ResNext-50 is lower than the performance of both ResNets.

\begin{table}
	\begin{minipage}{1.0\linewidth}
				\caption{Results of regression analyses (in terms of $R^2$) for $N=10,000$ test images in data sets shown in Table \ref{figdataaug} using the same data type for training, validation (model selection) and test.}
	\scalebox{0.80}{
		\begin{tabular}{ lccccccc }
			 & & \multicolumn{6}{c}{\textbf{ResNet-34}}
			 \\
			\multirow{8}{*}{{\rotatebox[origin=c]{90}{\textbf{\gls{psf} models}}}}    & \multicolumn{1}{c}{}
			&  \multicolumn{1}{c}{[syn]}
			&  \multicolumn{1}{c}{[poi]}
			& \multicolumn{1}{c}{[micr]}
			& \multicolumn{1}{c}{[micrsm]}
			& \multicolumn{1}{c}{[nat]}
			& \multicolumn{1}{c}{[micr-syn-poi]} \\
			\cmidrule[\heavyrulewidth]{3-8}

			  & Z-1 & 0.99 & 0.99 & 0.98 & 0.73 & 0.98 & 0.68\\
			  & Z-2 & 0.67 & 0.99 & 0.81 & 0.80 & 0.80 & 0.79\\
			  & Z-3 & 0.95 & 0.98 & 0.78 & 0.58& 0.84 &0.88\\
			  & G-1 & 0.99 & 0.99 & 0.98 & 0.92 & 0.92 & 0.99\\
			  & G-2 & 0.99 & 0.99 & 0.99 & 0.97 & 0.91 & 0.99\\
	\\
			 &	& \multicolumn{6}{c}{\textbf{ResNet-34-pretrained}}
				\\
	\multirow{8}{*}{{\rotatebox[origin=c]{90}{\textbf{\gls{psf} models}}}}    & \multicolumn{1}{r}{}
	&  \multicolumn{1}{c}{[syn]}
	&  \multicolumn{1}{c}{[poi]}
	& \multicolumn{1}{c}{[micr]}
	& \multicolumn{1}{c}{[micrsm]}
	& \multicolumn{1}{c}{[nat]}
	& \multicolumn{1}{c}{[micr-syn-poi]} \\
			\cmidrule[\heavyrulewidth]{3-8}
	  & Z-1 & 0.99 & 0.99 & 0.99 & 0.89 & 0.99 & 0.81\\
	  & Z-2 & 0.93 & 0.97 & 0.92 & 0.81 & 0.94 & 0.90\\
	  & Z-3 & 0.97 & 0.89 & 0.95 & 0.77 & 0.80 & 0.89\\
	  & G-1 & 0.94 & 0.99 & 0.98 & 0.99 & 0.94 & 0.99\\
	  & G-2 & 0.99 & 0.99 & 0.99 & 0.98 & 0.95 & 0.99\\
	\\
				 &	& \multicolumn{6}{c}{\textbf{ResNext-50}}
				 \\
	 \multirow{8}{*}{{\rotatebox[origin=c]{90}{\textbf{\gls{psf} models}}}}&\multicolumn{1}{r}{}
	&  \multicolumn{1}{c}{[syn]}
	&  \multicolumn{1}{c}{[poi]}
	& \multicolumn{1}{c}{[micr]}
	& \multicolumn{1}{c}{[micrsm]}
	& \multicolumn{1}{c}{[nat]}
	& \multicolumn{1}{c}{[micr-syn-poi]} \\
		\cmidrule[\heavyrulewidth]{3-8}
	  & Z-1 & 0.98 & 0.99 & 0.97 & 0.69 & 0.97 & 0.65\\
	  & Z-2 & 0.69 & 0.98 & 0.85 & 0.74 & 0.90 & 0.85\\
	  & Z-3 & 0.64 & 0.94 & 0.90 & 0.72 & 0.61 & 0.85\\
	  & G-1 & 0.99 & 0.99 & 0.97 & 0.80 & 0.91 & 0.98\\
	  & G-2 & 0.99 & 0.99 & 0.99 & 0.92 & 0.97 & 0.98\\
		
\end{tabular}
}
		\label{figcnnresults}
	\end{minipage}
\end{table}
\subsubsection{Characterization of \gls{cnn} regression performance when the type of training and test data set differ}
We assessed the robustness of our regression method when the system is tested on image types other than those it has been trained for.

\begin{framed}
	\textbf{Variables:} data set types for both training and test sets.
	\\
	\textbf{Fixed:} the network architecture (ResNet-34), the \gls{psf} model (Gaussian-2). The model is already trained and selected using an independent validation dataset.
	\\
	\textbf{Evaluation criterion:} $R^2$ between the degradation parameters used to generate the test input and the parameters recovered by the \gls{cnn}.
\end{framed}

Table \ref{fig:mismatchresults} gathers the regression performance obtained using a Gaussian-2 \gls{psf} model and the ResNet-34 network, with training and testing data sets of different types. The regression is robust to different train and test data set types ($R^2 < 0.90$) except when the \gls{cnn} is trained with [poi] and, to a lesser extent, with [syn]. Surprisingly, networks trained on natural ([nat]) images perform as well as networks trained on microscopy images.

\begin{table}
	\centering
	\begin{minipage}{1.0\linewidth}
		\centering
		\caption{Evaluation (in terms of $R^2$) for $N=10,000$ images input into a Resnet-34-pretrained network trained for regression of the Gaussian-2 \gls{psf} model parameters with different training set / test sets pairs. The learning rate and epoch were selected using results from a test set of $N=10,000$ separate images.}
		\scalebox{0.72}{
		\begin{tabular}{ lccccccc }
			 & & \multicolumn{6}{c}{\textbf{Test data sets}}
			\\
			\multirow{7}{*}{{\rotatebox[origin=c]{90}{\textbf{Training data sets}}}}    &  \multicolumn{1}{c}{}
								 	&  \multicolumn{1}{c}{[syn]}
								 	&  \multicolumn{1}{c}{[poi]}
								 	& \multicolumn{1}{c}{[micr]}
								 	& \multicolumn{1}{c}{[micrsm]}
								 	& \multicolumn{1}{c}{[nat]}
								 	& \multicolumn{1}{c}{[micr-syn-poi]} \\
			\cmidrule[\heavyrulewidth]{3-8}
			 & [syn] & 1 & 0.93 & 0.76 & 0.83 & 0.81 & 0.85\\
			 & [poi] & 0 & 1 & 0 & 0 & 0 & 0\\
			 & [micr] & 0.97 & 0.97 & 0.99 & 0.99 & 0.99 & 0.99\\
			 & [micrsm] & 0.97 & 0.95 & 0.95 & 0.98 & 0.89 & 0.97\\
			 & [nat] & 0.99 & 0.80 &  0.96 &0.93&0.96 & 0.96\\
			 & [micr-syn-poi] & 1 & 0.99 & 0.99  & 0.99 & 0.94 & 0.99\\
		\end{tabular}
		}
		
		\label{fig:mismatchresults}
	\end{minipage}
\end{table}

\subsection{Robustness of \gls{psf} regression against input degradation}
\label{sec:results_robustness}
Degradations on the input images are unavoidable in biological environments. Indeed, microscopes are often used for  a variety of sample types and preparations and are calibrated by different people. Settings such as illumination brightness, exposure time, and contrast frequently change or are operator-dependent. Furthermore, as described in Section \ref{sec:methods_dataset_generation}, low light and electronics induce noise in the acquired image. 
Since we aim at training a regression network that is not specific to any defined acquisition condition, we characterized the robustness of the neural network to extrinsic modifications of the image quality.

The list of handled degradations is the following:
\begin{itemize}
	\item  global illumination level, 
	\item  non-uniform illumination (e.g. caused by poorly adjusted K\"ohler illumination),
	\item  zero-mean Gaussian noise,
	\item  signal-dependent Poisson noise,
	\item  mixed Gaussian-Poisson noise.
\end{itemize}

\begin{framed}
	\textbf{Variables:} degradation strength, degradation type, three different \glspl{cnn}, two different \gls{psf} models (G-1, Z-1).
	\\
	\textbf{Fixed:} the \glspl{cnn} are already trained with [micr-syn-poi] and selected using an independent test dataset. The test data set is common to all modalities.
	\\
	\textbf{Evaluation criterion:} $R^2$ between the degradation parameters that were used to generate the test image and the parameters recovered by the CNN.
\end{framed}

\begin{figure}
	\begin{minipage}{1.0\linewidth}		
				\centering
		\includegraphics[width=\linewidth]{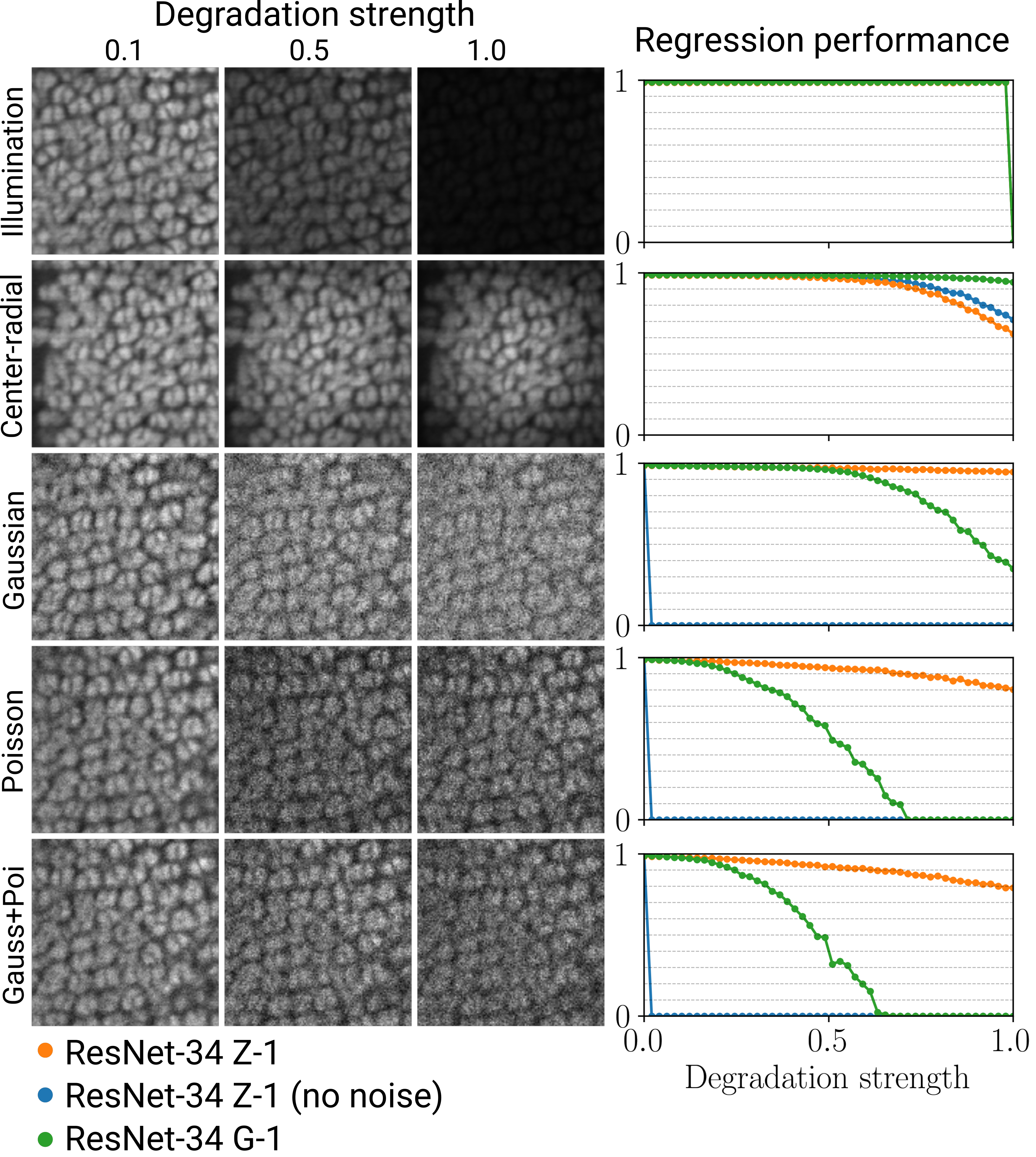}
		\caption{Regression performance of \glspl{cnn} architectures trained on [micr] and evaluated on the [micr] test set with various types of degradation and variable strength. The regression performance is shown in terms of $R^2$ using $N=10,000$ images. We trained as well a \gls{cnn} without adding noise in the training set (blue). All three networks exhibits the same performance for the illumination degradation (top right).}
		\label{fig:degradation_results}
	\end{minipage}
\end{figure}

Fig. \ref{fig:degradation_results} summarizes the performance of the \gls{cnn} networks as a function of degradation strength. All networks are robust to partial or full brightness changes in the input image. The addition of Gaussian noise to the input results in a slow and linear decay in performance, whereas the application of Poisson noise to the data decrease the performance much faster as the noise strength increases.
Using \glspl{cnn} trained for regression of Zernike polynomial \gls{psf} model parameters, the regression performance is decaying linearly as a function of the amount of noise we apply in the input picture. For Gaussian models, the parameter estimation usually breaks with less noise than with the Zernike polynomial model.
Without any degradation, networks regressing Zernike polynomials are less accurate than networks for Gaussian \gls{psf} models, but they appear to be more robust when the input is noisy. Indeed, with a very strong (strength of $0.5$ in Fig.~\ref{fig:degradation_results}) Gaussian and Poisson noise, Zernike-1 \glspl{cnn} dropped from $R^2=0.95$ to $R^2=0.85$, as opposed to the Gaussian-1 \glspl{cnn}, which dropped from $R^2=0.99$ to less than $R^2 = 0.60$. Surprisingly, contrary to the findings in a recent benchmark \cite{hendrycks_benchmarking_2019}, we found that \glspl{cnn} based on ResNeXt performed worse than their ResNet counterparts. Finally, we trained new models without adding synthetic noise to the training dataset (see Eq. (\ref{eq:methods_degradation})). Performance of these networks was the same as their counterparts for the illumination degradations, but dropped to $R^2 = 0$ when the test images contained even only moderate Poisson and Gaussian noise.

\subsection{Application 1: spatially-variant blind deconvolution}
We next wanted to verify that the parameters recovered by the \gls{cnn} were producing \glspl{psf} that are sufficiently accurate to be usable to enhance the details in the image, despite not being specifically measured. To this end, we devised a deconvolution experiment to compare images deconvolved by our method with those obtained by other blind deconvolution techniques. As test input, we used $256 \times 256$ pixels image patches from the [micr] data set (see Section \ref{sec:methods_dataset_generation}). Using Eq. (\ref{eq:methods_degradation}), we degraded each quadrant of the input image with a specific, randomly-generated $127 \times 127$ pixels \glspl{psf} using parameters $\boldsymbol{a}(\bm{s})$ drawn from a uniform random distribution allowing us to systematically explore the parameter space. We subsequently inferred a \glspl{psf} map via the \gls{cnn} from the blurry image as described in Section \ref{sec:methods_psfmapping}. Finally, we deconvolved the image by applying the method described in Section \ref{sec:application1}. In the experiment, we set $\varphi^{(m)}$ as a bilinear interpolating function and, similarly to \cite{dey_richardsonlucy_2006}, $\lambda_{TV} = 0.1$. Since using FFT-based calculations implies that the \gls{psf} is circulant, we took into account field extension to prevent spatial aliasing. We fixed the number $I$ of \gls{rl} iterations to $20$.
\\
We assessed the reconstruction quality by computing the \gls{snr} and \gls{ssim} \cite{wang_multiscale_2004}. We compared the deconvolution results to spatially-invariant blind deconvolution techniques \cite{kotera_blind_2013}, \cite{dong_blind_2017} and \cite{jin_normalized_2018}, and the spatially-variant method from \cite{whyte_deblurring_2014}. In the latter cases, we used the estimated PSF in the \gls{tvrl} algorithm with the same number of iterations and $\lambda_{TV}$. Since the estimation of a full PSF by these methods would take more than 20 minutes per sample, we constrained the support of the \gls{psf} to $31 \times 31$ pixels.
We computed the scores by taking the difference between the ``ground truth" \gls{snr} and \gls{ssim} of images deconvolved using the \glspl{psf} actually used to degrade the images, and the deconvolution results using \glspl{psf} regressed with the \gls{cnn} or other \gls{bd} techniques. Theses values are therefore reported as $\Delta \text{SNR}$ and $\Delta \text{SSIM}$.
\\
Finally, in order to recover details lost due to the aberrations of actual microscope objectives (such as out-of-focus blur and astigmatism), we acquired different fixed samples (HeLa cells actin (Alexa Fluor 635) and HeLa cells anti-$\alpha$-catenin (Alexa Fluor 488) with a 10$\times$/0.3 air objective, \textit{Convallaria majalis} bulb autofluorescence with a 20$\times$/0.7 air objective) both in focus and slightly out of focus. Then, starting from a $256 \times 256$ patch of the out-of-focus picture only, we sought to retrieve a sharper picture containing the details of the in-focus picture. We compared qualitatively the in-focus image, the out-of-focus image, our method with four \glspl{psf} detected with a $128 \times 128$ stride, \cite{kotera_blind_2013}, \cite{whyte_deblurring_2014}, and the imaged obtained via a ``sharpen'' high-pass filter.

\begin{framed}
	\textbf{Variables:} network architecture, two different \gls{psf} models (Z-1 and G-1) for the degradation and detection parts.
	\\
	\textbf{Fixed:} \glspl{cnn} are already trained, the test data set [micr] is fixed.
	\\
	\textbf{Evaluation criterion:} difference of \gls{snr} and \gls{ssim} between the ground truth image and the deconvolved image.
\end{framed}

\begin{figure}
	\begin{minipage}{1.0\linewidth}		
		\centering
		\includegraphics[width=\linewidth]{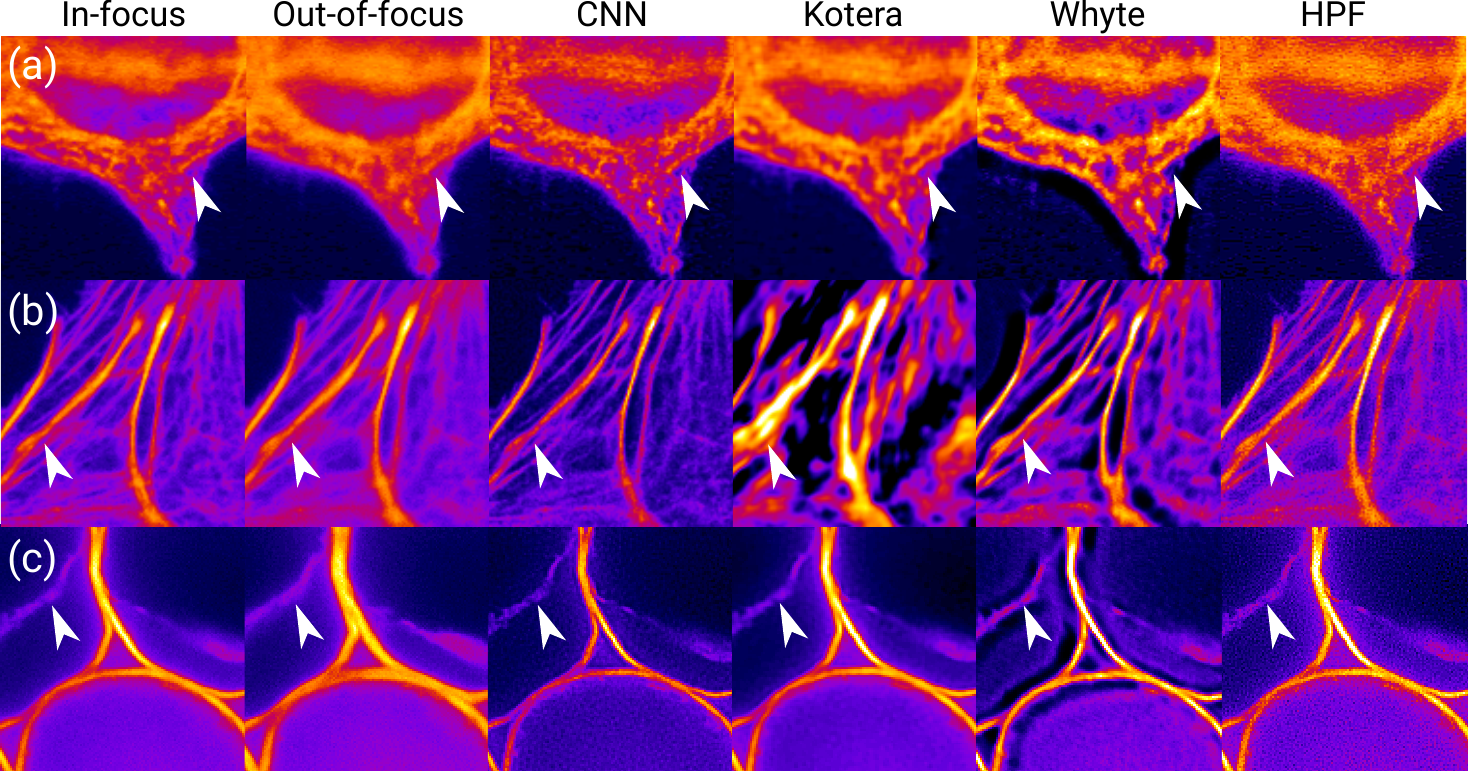}
		\caption{We acquired different fixed samples ((a) HeLa cells actin (Alexa Fluor 635) and (b) HeLa cells anti-$\alpha$-catenin (Alexa Fluor 488) with a 10$\times$/0.3 air objective, (c) \textit{Convallaria majalis} bulb autofluorescence with a 20$\times$/0.7 air objective) both in-focus and slightly out-of-focus. Then, starting from a out-of-focus $256 \times 256$ patch only, we seek to retrieve a sharper picture containing the details of the in-focus picture. We compared qualitative results of (from left to right), the in-focus image, the out-of-focus image, the proposed method (\gls{cnn}) with four \glspl{psf} detected with a $128 \times 128$ stride, \cite{kotera_blind_2013}, \cite{whyte_deblurring_2014}, and a "sharpen'' high-pass filter. Arrow edges indicate regions with features of particular interest.}
		\label{fig:comparison_results}
	\end{minipage}
\end{figure}

Results in Table \ref{fig:deconvresults} indicate an average improvement  of both \gls{snr} ($1.88$ dB) and \gls{ssim} ($0.09$) of our spatially variant \gls{bd}. In comparison to spatially-invariant \gls{bd} and other spatially-variant \gls{bd} techniques improves the image by $1.55$ dB \gls{snr} and 0.08 \gls{ssim}. Deconvolution results are equivalent when the degradation and detection models are mismatched.
The qualitative results shown in Fig.~\ref{fig:comparison_results} highlight the stability of our method, which improves the degraded image to a detail level similar or better than the one of the in-focus image. Using the algorithm of Kotera et al. \cite{kotera_blind_2013}, the blurry features are well recovered, but the images have less detail. Furthermore, this algorithm converges to an aberrated image (Fig.~\ref{fig:comparison_results} (b)) when the image contains long filaments. The method from Whyte et al. \cite{whyte_deblurring_2014} enhances the contrast of the blurry image, however, it creates hallucinations near edges, which were not part of the original image. Finally, the high-pass filter, as expected, enhances both high-frequency features and noise.

\begin{table*}
			\caption{Image deconvolution benchmark with 350 input images ($256 \times 256$) from the test set [micr] for each experiment. We computed the difference in \gls{snr} and \gls{ssim} between the ground truth image and the deconvolved image. We compared the following methods: \gls{tvrl} with four known \glspl{psf} (gt), \gls{tvrl} with four \glspl{psf} estimated by the proposed \gls{cnn} (CNN$_{\text{blind}}$),  Kotera et al., 2013 \cite{kotera_blind_2013}, Whyte et al., 2014 \cite{whyte_deblurring_2014}, Dong et al. \cite{dong_blind_2017} and Jin et al. \cite{jin_normalized_2018}. The \gls{cnn} is a ResNet-34 trained on data set [micr].}

	\begin{minipage}{1.0\linewidth}
					\centering
			\scalebox{0.90}{
		\begin{tabularx}{1.11\linewidth}{ Y Y c c c c c c }

			\multicolumn{2}{c}{\makecell{blurry \\ \includegraphics[width=2.2cm]{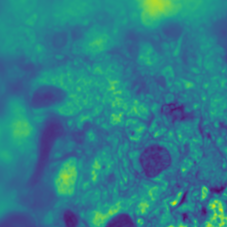}}}  &\makecell{gt\\\includegraphics[width=2.2cm]{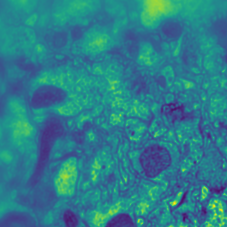}} & \makecell{CNN$_{\text{blind}}$\\\includegraphics[width=2.2cm]{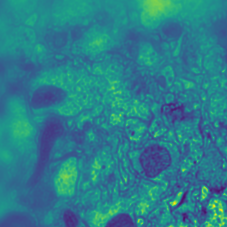}}& \makecell{Kotera \cite{kotera_blind_2013}\\\includegraphics[width=2.2cm]{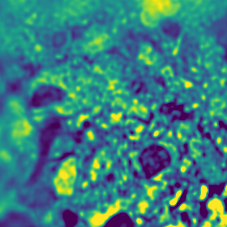}} & \makecell{Whyte \cite{whyte_deblurring_2014}\\\includegraphics[width=2.2cm]{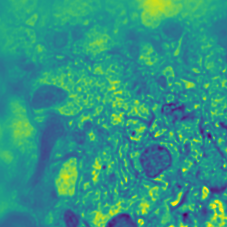}}& \makecell{Dong \cite{dong_blind_2017}\\\includegraphics[width=2.2cm]{deconvolved_wyss}}& \makecell{Jin \cite{jin_normalized_2018}\\\includegraphics[width=2.2cm]{deconvolved_wyss}}\\
			\toprule
			\multicolumn{1}{Y}{{Degradation}}
			&  \multicolumn{1}{Y}{{Detection}}
			&  \multicolumn{1}{c}{{$\Delta\text{SNR}_\text{gt}$}}
			&  \multicolumn{1}{c}{{$\Delta\text{SNR}_\text{CNN-blind}$}}
			& \multicolumn{1}{c}{{$\Delta\text{SNR}_\text{\cite{kotera_blind_2013}}$}}
			& \multicolumn{1}{c}{{$\Delta\text{SNR}_\text{\cite{whyte_deblurring_2014}}$}} 
			& \multicolumn{1}{c}{{$\Delta\text{SNR}_\text{\cite{dong_blind_2017}}$}}
			& \multicolumn{1}{c}{{$\Delta\text{SNR}_\text{\cite{jin_normalized_2018}}$}}\\
			\cmidrule(){1-8}
			Z-1& Z-1 &$ 2.14 \pm 0.71$ & $\mathbf{1.65} \pm 0.62$ & $-0.81 \pm 0.71$ & $0.12 \pm 0.31$ & $0.61 \pm 0.58$ & $1.34 \pm 0.62$\\
			Z-1& G-1 &$ 2.14 \pm 0.71$ & $\mathbf{1.72} \pm 0.71$ & $-0.61 \pm 0.67$ & $0.02 \pm 0.31$ &$0.83 \pm 0.59$ &$1.48 \pm 0.67$\\
			G-1& Z-1 &$2.24 \pm 0.73$ & $\mathbf{2.02} \pm 1.04$ &  $-1.20 \pm 0.97$ & $-0.01 \pm 0.55$& $0.89 \pm 0.74$& $1.78 \pm 0.76$\\
			G-1& G-1 &$2.24 \pm 0.73$ & $\mathbf{2.11}\pm 0.72$ & $-1.25 \pm 1.03$ & $-0.31 \pm 0.87$ & $0.68 \pm 0.9$ & $1.58 \pm 0.70$\\
			\toprule
			\multicolumn{1}{Y}{{}}
			&  \multicolumn{1}{Y}{{}}
			&  \multicolumn{1}{c}{{$\Delta\text{SSIM}_\text{gt}$}}
			&  \multicolumn{1}{c}{{$\Delta\text{SSIM}_\text{CNN-blind}$}}
			& \multicolumn{1}{c}{{$\Delta\text{SSIM}_\text{\cite{kotera_blind_2013}}$}}
			& \multicolumn{1}{c}{{$\Delta\text{SSIM}_\text{\cite{whyte_deblurring_2014}}$}} 
			& \multicolumn{1}{c}{{$\Delta\text{SSIM}_\text{\cite{dong_blind_2017}}$}}
			& \multicolumn{1}{c}{{$\Delta\text{SSIM}_\text{\cite{jin_normalized_2018}}$}}\\
			\cmidrule(){1-8}
			Z-1 & Z-1 & $ 0.10 \pm 0.03$ & $\mathbf{0.10} \pm 0.04$  & $0.05 \pm 0.04$ & $0.02 \pm 0.02$&$0.09 \pm 0.04$ & $0.09 \pm 0.04$\\
			Z-1& G-1 & $ 0.10 \pm 0.03$ & $\mathbf{0.09} \pm 0.03$  & $0.03 \pm 0.05$ & $0.02 \pm 0.01$ & $0.08 \pm 0.04$& $0.08 \pm 0.03$\\
			G-1& Z-1 &$0.08 \pm 0.03$ & $0.08 \pm 0.06$ & $0.01 \pm 0.05$ & $0.01 \pm 0.03$&$0.07 \pm 0.04$ &$\mathbf{0.08} \pm 0.03$\\
			G-1& G-1 &$0.08 \pm 0.03$ & $0.08 \pm 0.04$ & $0.01 \pm 0.06$ & $0.03 \pm 0.03$& $0.07 \pm 0.04$ & $0.08 \pm 0.04$\\
			\bottomrule
		\end{tabularx}
	}

		\label{fig:deconvresults}
	\end{minipage}
\end{table*}

\subsection{Application 2: depth from focus using astigmatism}
The \gls{psf} parameter describing the local blur that we obtained from the image in Section \ref{sec:methods_psfmapping} lacks information about the relative direction of the object from the focal plane. To infer the local axial distance map $z(\bm{s})$ in the sample, we applied the method described in Section \ref{sec:application2} using the astigmatism created by a cylindrical lens. As an imaging sample, we used a grid of $\SI{200}{\micro\metre} \times \SI{200}{\micro\metre}$ squares which we laser-printed on a transparent plastic foil. We placed the grid towards the focal plane and tilted it by $3^{\circ}$, $6^{\circ}$, or $10^{\circ}$, so that the in-focus position was in the middle of the field of view (Fig. \ref{fig:application2} (a)). With a field of view of $\SI{655}{\micro\metre} \times \SI{655}{\micro\metre}$, such a rotation yielded depth ranges of $\SI{48.2}{\micro\metre}$, $\SI{94.7}{\micro\metre}$, or $\SI{159.9}{\micro\metre}$, respectively.

We were able to retrieve the local Zernike coefficient parameters of focus ($a_1$), cylinder ($a_2$), and axis ($a_3$). We inferred the depth map $z(\bm{s})$ using Eq. (\ref{eq:application2}) for $N=30$ acquired images in total (see Fig. \ref{fig:application2} (e)).

\begin{framed}
	\textbf{Variables:} input images of surfaces with a varying tilt angle from the focal plane.
	\\
	\textbf{Fixed:} the network architecture (ResNet-34-pretrained for regression of Zernike parameters with astigmatism). \glspl{cnn} are already trained, the test data set is fixed.
	\\
	\textbf{Evaluation criterion:} $\ell_1$-error between the actual axial position of the surface and the position extracted from the image.
\end{framed}

Using a \gls{cnn} ResNet-34 trained with [micr], we obtained a correlation coefficient ($R^2$) between the average slope and a line fit of more than $0.96$. From this line fit we calibrated the system to spatial units so that we could build a depth map. The method was accurate with an absolute average $\ell_1$-error of $\SI{1.81 \pm 1.39}{\micro\meter}$ in depth (corresponding to a $1.61 \pm 1.23 \%$ of the relative depth boundaries), obtained by comparing the error between the known position of the object in depth and the calibrated distance. Results in Table \ref{fig:application2table} reveal that the relative error increases as the maximum depth of the object increases. Depth estimation is thus more precise around the focal plane.
\begin{table}
	\begin{minipage}{1.0\linewidth}
		\centering
		\caption{Analysis of the precision of the depth recovery of a plane using controlled astigmatic aberrations by use of cylindrical lenses. The test images are acquisitions of $N=30$ printed grids tilted 3$^{\circ}$, 6$^{\circ}$ and 10$^{\circ}$. The network is a ResNet-34 trained on data set [micr].}
								\scalebox{0.90}{
			\begin{tabularx}{1.11\textwidth}{  c c Y Y }
				\toprule
				  \multicolumn{1}{c}{Tilt angle}
				&  \multicolumn{1}{c}{$R^2$}
				& \multicolumn{1}{Y}{Absolute error} 
				& \multicolumn{1}{Y}{Relative error} \\
				\cmidrule{1-4}
				10$^{\circ}$ & 0.967 & $\SI{3.50 \pm 2.62}{\micro\meter}$ & $2.18 \pm 1.64 \%$\\
				6$^{\circ}$ & 0.988 & $\SI{1.31 \pm 1.15}{\micro\meter}$ & ${1.38 \pm 1.21}\%$\\
				3$^{\circ}$ & 0.989 & $\SI{0.61 \pm 0.41}{\micro\meter}$ & ${1.26 \pm 0.85}\%$\\
				\bottomrule
			\end{tabularx}
		}
		\label{fig:application2table}
	\end{minipage}
\end{table}

\section{Discussion}
\label{sec:discussion}
Hereafter, we discuss the results of the experiments described in Section \ref{sec:experiments}.
\subsection{Characterization of the \gls{cnn} regression performance}
Many regression accuracies are above $R^2 = 0.90$ in Table \ref{figcnnresults}, which shows that our neural networks can accurately regress \gls{psf} model parameters (in particular, when images are textured). The recovered parameters can be used to generate synthetic \glspl{psf} that are similar to the ones that degraded the image.
\\
Our network is most accurate when applied to images of the same type as the ones used for training and the performance scales with the size of the training set. Therefore, the more data we gather, the more precise and robust the predictions are. However, adding synthetic training data to augment a natural images data set does not increase the efficacy of \gls{psf} estimations. The network fails to predict the \gls{psf} parameters if the network is trained with a very narrow type of data and more variety increases generalization potential. Nevertheless, when learning from a data set of images containing texture, even if different from the test set type, the model remains as accurate as when training the data set using microscopy images. This suggests that one could avoid the need to gather costly microscopy ground-truth data, or that it might be possible to learn from images of other microscope types (e.g. confocal microscopes) and use the trained models with wide field microscopy images. Furthermore, the high correlation score ($R^2 > 0.8$) obtained for [micr] test images using networks trained with [nat] suggests that the networks did not undergo overfitting and were able to generalize on other data types.
\\
In comparison to ResNet-34, ResNeXt-50 requires a larger number of images to be accurate since the regression accuracy drops drastically (from $R^2 = 0.97$ to $R^2 = 0.79$) using a smaller data set. It is consistent with the general idea that the amount of  training data must scale with the depth of the network to be able to generalize well.
\\
Networks trained for Gaussian \glspl{psf} estimate parameters with a better accuracy than networks trained to find Zernike polynomials parameters. This could be explained by the fact that although the Zernike polynomial parameters are independent when describing the pupil function, they can compensate each other when forming the \gls{psf} (which is obtained by a non-linear operation on the pupil function, see Eq.~(\ref{eq:pdffromzernike})). 
\\
Using an NVIDIA GeForce GTX 1080 GPU, the estimation of the \gls{psf} parameters of a $1024 \times 1024$ px image with 64 \glspl{psf} takes around $\SI{5.9 \pm 0.1}{\milli\second}$, which is in the same scale as the usual camera exposure time. This suggests that real-time applications in a microscope could be feasible.
\\
While transfer learning (i.e. networks trained with ImageNet for image detection prior to training) does not help to improve the final accuracy of the regression task because the network was already able to learn from the original data set in a reasonable amount of time, training networks by starting from pre-trained models tends to speed up convergence during learning.

\subsection{Robustness analysis against input degradation}

We noticed that all our models are invariant to changes in illumination, certainly due to inherent normalization steps in the \gls{cnn} architecture, and are overall robust to small to medium amounts of noise. However, when the signal-to-noise ratio strongly decreases, the correlation coefficient tends to decrease as well.

Additionally, we observed that, in comparison to ResNet, \glspl{cnn} based on ResNeXt perform generally worse when noise is applied. This result contradicts observations reported in a recent benchmark \cite{hendrycks_benchmarking_2019}, where ResNext is more robust than ResNet to Gaussian noise. However, this publication scores the network's accuracy for a classification task into image type categories, which is a different application than our regression of aberrations and might explain the discrepancy.

We found the synthetic degradations we added in the training set (Eq. (\ref{eq:methods_degradation})) to be a necessary step to achieve robustness to noise. Indeed, the performance dropped when this step was omitted.

\subsection{Application 1: spatially-variant blind deconvolution}

Using the spatially-variant \gls{psf} map inferred from the \gls{psf} output, we have been able to reconstruct details in a degraded image without any prior information on the image content or the optical system. Given that our method does not require adjusting parameters or experimentally measuring a \gls{psf} (which is labor intensive), it leads to results faster than non-blind deconvolution methods.

We noticed that, because we use a constrained \gls{psf} model, our deconvolution method does not suffer from drawbacks sometimes associated to other deconvolution and super-resolution methods. For example, techniques based on \gls{map} optimization sometimes converge to exotic forms of \glspl{psf} that are not consistent with the physics of optics, causing image deformation or loss of features \cite{perrone_total_2014}. We could illustrate this by the example in Fig.\ref{fig:comparison_results} (b) for the method from Kotera et al. \cite{kotera_blind_2013}, which diverges when directed filaments are shown to the algorithm and creates artifacts.

Similarly, the use of denoising \glspl{cnn} for image enhancement can lead to phantom details that could falsify underlying biological features, or discard high-frequency features that are mistaken for noise \cite{liu_when_2018}. When using constrained \gls{psf} models such as the ones we use, deconvolution algorithms such as \gls{rl} will still produce a reasonable image even if the predicted \gls{psf} is not exactly matching the \gls{psf} corresponding to the blur. This is likely due to the inherent constraint of a model with a small number of parameters that enforces the shape of the \gls{psf}. The outcome is a higher average \gls{snr} and \gls{ssim} of the reconstruction using our method compared to other \gls{bd} algorithms. Another advantage of using constrained \gls{psf} models is that they can model \glspl{psf} with very large supports (sizes). Classical \gls{bd} only allows for a smaller support, as using more pixels creates higher complexity. Nevertheless, our models are currently unlikely suitable for some types of degradation, where \gls{bd} methods were successfully applied, such as for compensating for motion blur with rotation \cite{shan_high-quality_2008}.

Deconvolution results are equivalently efficient both in terms of \gls{snr} and \gls{ssim} when the degradation and detection models are not similar (e.g. degradation using a Gaussian \gls{psf} and estimation of the \gls{psf} using a Zernike model). This particular point is relevant since it confirms the robustness of the image enhancement process when there is a mismatch between the degradation \gls{psf} that we want to model (i.e. the optical system \gls{psf}) and the model itself.

Finally, we observed that methods \cite{whyte_deblurring_2014}, \cite{dong_blind_2017} and \cite{jin_normalized_2018}, due to their multiscale optimization approach, were considerably slower than the one we propose, taking up to 4 minutes to deblur a $256 \times 256$ image, whereas our method takes less than 3 seconds using the same machine to both estimate the PSF and perform \gls{tvrl}. The difference in run times can be explained by our GPU implementation, but as well because of the inherent nature of traditional optimization algorithms that alternate kernel and image estimation, which limits the parallelizability of the calculations.

\subsection{Application 2: depth from focus using astigmatism}
We showed that the focus parameter of the \gls{psf} models is a function of the distance between the sample and the focal plane, and that the sign of the axial distance could be recovered from higher Zernike coefficients when using engineered \glspl{psf}. Furthermore, we found that the depth function at any point in the image could be obtained from an affine function that combines two Zernike coefficients.

We recovered the relative depth of both sides of the printed grid in real microscopy acquisitions, which extends the idea of \gls{psf} engineering introduced in \cite{huang_three-dimensional_2008} for point-like structures, to work for fully textured images. Using a textured plane image, we have been able to recover the depth to up to \SI{160}{\micro\meter}. However, this accuracy decreases when the imaged objects lack texture. Using other shapes of engineered \glspl{psf} (e.g. quadripoles \cite{aristov_zola-3d_2018}) could potentially lead to improved depth accuracy.

Since use of an engineered \gls{psf} degrades the image, simultaneous depth retrieval and high-resolution might best be carried out by splitting the acquisition line (e.g. with a beam splitter) to record the image on one side and a \gls{psf} engineered image on the other. 

Depth retrieval is also possible using the parameters of a Gaussian-2 \gls{psf} model, but the precision is improved using the \gls{psf} model based on Zernike polynomials.

\section{Conclusion}
\label{sec:conclusion}
In this work, we have shown that \glspl{cnn}, in particular residual networks, can be used to extract local blur characteristics from microscopy images in the form of parameters of a \gls{psf} model with only minimal knowledge about the optical setup. Our system is robust to signal perturbations and does not need to be trained specifically on images of the target imaging system. This flexibility allows the user to perform, without taking measurements beyond the images of interest, a wide range of tasks in microscopy image processing, including deblurring or obtaining a depth map. 

Our method opens up the possibility to improve the resolution of non-flat objects with minimal \textit{a priori} knowledge of the optical setup and obtain its tridimensional shape in a single shot.

\appendix{}
Software sources, documentation, trained models, and the acquired test dataset (Fig.~\ref{fig:comparison_results}) will be made available upon publication. We thank Arne Seitz and José Artacho from EPFL BIOP for providing the fixed sample slides, reused from experiments that were approved by the EPFL Ethics Committee.

\section{References}
\label{sec:references}
\begin{spacing}{0.9}
\printbibliography[]
\end{spacing}

\end{document}